\documentclass[prb,twocolumn,floatfix,showpacs,preprintnumbers,superscriptaddress,amsmath,amssymb]{revtex4}

\usepackage{graphicx}

\begin{document}

\title{Structure and conductance histogram of atomic-sized Au contacts}

\author{M. Dreher}
\affiliation{Physics Department, University of Konstanz, 78457 Konstanz, Germany}

\author {F. Pauly} 
\affiliation{Institut f\"ur Theoretische Festk\"orperphysik, University of Karlsruhe,
76128 Karlsruhe, Germany}

\author {J. Heurich} 
\affiliation{Institut f\"ur Theoretische Festk\"orperphysik, University of Karlsruhe,
76128 Karlsruhe, Germany}

\author {J.C. Cuevas}
\affiliation{Institut f\"ur Theoretische Festk\"orperphysik, University of Karlsruhe,
76128 Karlsruhe, Germany}

\author{E. Scheer}
\affiliation{Physics Department, University of Konstanz, 78457 Konstanz, Germany}

\author{P. Nielaba}
\affiliation{Physics Department, University of Konstanz, 78457 Konstanz, Germany}

\date{\today}

\begin{abstract}
Many experiments have shown that the conductance histograms of metallic atomic-sized
contacts exhibit a peak structure, which is characteristic of the corresponding
material. The origin of these peaks still remains as an open problem. In order to shed
some light on this issue, we present a theoretical analysis of the conductance histograms
of Au atomic contacts. We have combined classical molecular dynamics simulations of the
breaking of nanocontacts with conductance calculations based on a tight-binding model. 
This combination gives us access to crucial information such as contact geometries,
forces, minimum cross-section, total conductance and transmission coefficients of the
individual conduction channels. The ensemble of our results suggests that the low 
temperature Au conductance histograms are a consequence of a subtle interplay between
mechanical and electrical properties of these nanocontacts. At variance with
other suggestions in the literature, our results indicate that the peaks in the Au conductance
histograms are not a simple consequence of conductance quantization or the existence
of exceptionally stable radii. We show that the main peak in the histogram close
to one quantum of conductance is due to the formation of single-atom contacts and chains of 
gold atoms. Moreover, we present a detailed comparison with experimental results on Au atomic
contacts where the individual channel transmissions have been determined.
\end{abstract}


\maketitle

\section{Introduction}
The metallic nanowires fabricated by means of the scanning-tunneling microscope 
and break-junction techniques have turned out to be an unique playground where
to test concepts on electronic transport at the atomic scale~\cite{Agrait2003}.
The great activity in this field during the past decade was triggered by the 
observation of a step-like evolution of the conductance during the formation of
the nanowires~\cite{Muller1992,Agrait1993}. Immediately, the connection with 
conductance quantization (CQ) observed in two-dimensional electron-gas devices~\cite{CQ}
was discussed, where every conduction channel has a perfect transparency. In order 
to investigate in detail CQ, different authors introduced
conductance histograms, constructed from a large number of individual conductance
curves~\cite{Olesen1995,Krans1995,Gai1996}, i.e. the evolution of the conductance 
with the elongation. These demonstrated that for metals like gold and sodium the 
conductance has a certain preference to adopt multiples of $G_0=2e^2/h$, the quantum
of conductance. However, for a large variety of metals, the peaks in the histograms 
do not appear at multiples of $G_0$ (for a detailed discussion of the conductance 
histograms, see section V.D in Ref.~[\onlinecite{Agrait2003}]). For instance, it has been 
shown~\cite{Scheer1997} that an Al contact sustains up to three conduction channels
in the last conductance plateau, although the typical conductance is close to $1~G_0$. 
The Al histogram exhibits a series of peaks~\cite{Yanson1997}, which in view of the 
results of Ref.~[\onlinecite{Scheer1997}] cannot be interpreted as CQ. These results 
showed that one cannot simply interpret the peaks close to multiples of $G_0$ in the 
histograms as evidence for CQ~\cite{Yanson1997}. The results of Ref.~[\onlinecite{Scheer1997}]
were explained on the basis of a tight-binding model~\cite{Cuevas1998a,Cuevas1998b}. 
From this work the picture that emerged was that the number of conduction channels is
limited by the valence orbitals of the corresponding metal, and the channels are 
in general not fully transparent. This view was confirmed
experimentally by a systematic study for various metals~\cite{Scheer1998}.
In spite of the success of this atomic picture, the origin of the peaks in the 
conductance histograms remains unclear, and the main goal of this work is to 
shed some light on this issue. 

It is obvious that the mechanical properties of the nanowires can play an
important role in the shape of a conductance histogram. This was pointed out in early
molecular-dynamics simulations~\cite{Landman1990,Sutton1990,Todorov1993}, which 
suggested that the steps in the conductance during the formation of the nanowires are
due to atomic rearrangements. Clear experimental evidence of this fact was reported by Rubio 
\emph{et al.}~\cite{Rubio1996}, who combined conductance and force measurements to
show that the jumps in the conductance are associated with distinct jumps in the force.
A striking example of the interplay between mechanical and electrical properties can
be found in the work of Yanson \emph{et al.}~\cite{Yanson1999,Yanson2001}. These authors 
observed a series of peaks in the histograms of alkali metals at relative high temperatures,
which were convincingly interpreted as the existence of exceptionally stable contact 
diameters due to electronic and atomic shell effects. This confirmed the idea that the
quantum modes in nanowires not only determine the conductance but also play an important 
role in the cohesive energy~\cite{Stafford1997,Ruitenbeek1997,Yannouleas1997}. More recently, 
Hasmy \emph{et al.}~\cite{Hasmy2001} have calculated histograms of the minimum cross-section
for Al contacts using molecular dynamics. At low temperatures they obtained
peaks at multiples of the cross-section of a single atom, which led them to a
new interpretation of the conductance histogram peaks based on preferential geometrical
arrangements of nanocontact necks. It is also important to mention that very
recently the room temperature conductance histograms of Au and Al have been interpreted
as an evidence of electronic and atomic shell effect in these 
metals~\cite{Medina2003,Mares2004}.

So far, several publications have analysed the evolution of metallic
nanowires
in a stretching event~\cite{tbdftmd,Ugarte2004}.
The theoretical analysis of conductance histograms, however, has only been carried out 
within the framework of free-electron models by choosing
particular nanowire dynamics~\cite{Torres1996}. Some authors have combined classical 
molecular-dynamics simulations with atomistic calculations of the 
conductance~\cite{Todorov1993,Bratkovsky1995,Todorov1996,Mehrez1997,Brandbyge1997,
Sorensen1998,Buldum1998,Nakamura1999} to analyze single realizations of the contact.
The existence of some exceptionally stable structures and the correlation of conductance
histograms and atomic structure were previously published for gold using the Wulff-construction
method and the H\"uckel approximation for the conductance~\cite{Ugarte2000,Ugarte2003}.
However, there are no full atomistic studies of the conductance histograms.

In this work we report for the first time a full atomistic study of
the conductance histograms of gold atomic contacts, and we present an extensive
comparison with experimental results. For this purpose, we have 
combined atomistic calculations of the conductance within a tight-binding approximation
with realistic molecular dynamics simulations of contact geometries. 
This combination allows us to obtain detailed information on the mechanical and transport 
properties such as contact geometries, forces, minimum cross-section, total
conductance, the number and the evolution of individual transmission coefficients.  
We have chosen the element gold for two reasons. First, gold histograms have been widely 
investigated~\cite{Olesen1995,Krans1995,Muller1996,Costa-Kramer1997,Hansen1997,
Li1998,Ludoph2000}. One typically observes a pronounced peak at $1~G_0$, and much
smaller peaks close to $2~G_0$ and $3~G_0$, although the histograms slightly depend on
details such as temperature, voltage bias and vacuum conditions. Second, the experimental
analysis of the individual
conduction channels using the technique introduced in Ref.~[\onlinecite{Scheer1997}] 
is available~\cite{Scheer1998,Scheer2001,Scheer2002,Rubio2003}. This analysis shows that 
conduction channels disappear one by one as the contact is elongated, and in the
last plateau, which corresponds to either a single-atom contact or a 
chain~\cite{Ohnishi1998,Yanson1998}, the conductance is dominated by a single channel.

Our results show that the histograms of the minimum cross-section at low temperatures 
(4 K) exhibit peaks, suggesting the existence of some exceptionally stable structures.
From a semi-classical point of view, one would then expect these peaks to be reflected
as peaks in the conductance histograms. However, our results indicate that this is
in general not the case. Indeed, only in the last stage of the stretching process we 
obtain a large peak close to $1~G_0$, which is correlated to the formation of a single-atom
contact or a chain of atoms. Our channel analysis shows that the conduction modes disappear
one by one and the last plateau is dominated by a single channel, in agreement with the 
experimental results. In short, the ensemble of our results shows that the low-temperature
conductance histograms of gold are a consequence of the interplay between the 
mechanical and the electrical properties of these atomic contacts.

The rest of the paper is organized as follows. In section II we explain the
technical details of our molecular dynamics simulations and conductance calculations.
Section III is devoted to the analysis of some examples of the breaking of the
contacts, where in particular we show the evolution of the single channels and illustrate
the typical geometries formed in the last stages: single-atom contacts and chains of atoms. 
In section IV we present new experimental results of the determination of the channel 
transmissions in Au contacts and we compare these results with our theoretical simulations. 
In section V we discuss our results for the histograms of the minimum cross-section. 
The corresponding results for the conductance histogram are presented in 
section VI and their correlation with the minimum cross-section histograms is discussed 
in detail. Finally, in section VII we present a brief summary of the main results and 
conclusions.

\section{Description of the theoretical model}

As mentioned in the introduction, in order to study the histograms of gold nanowires
we have performed classical molecular dynamics (MD) simulations to investigate the
formation of the contacts, and we have used a tight-binding model to calculate
the conductance during the breaking of the atomic necks. It is worth stressing that
the resulting numerical effort is considerable~\cite{numEffort}. In what follows 
we explain the details of both theoretical techniques.

\subsection{Molecular dynamics simulations}

We analyze the breaking of Au nanocontacts by means of classical MD simulations.
The forces and energies are calculated using semi-empirical potentials derived from
the effective medium theory~\cite{Jacobsen1996}, which successfully describe experimental
bulk and surface properties \cite{Stoltze1994} and have also been successfully used for simulating 
nanowires by Jacobsen et. al.~\cite{Brandbyge1997,Sorensen1998,Nakamura1999,Rubio2002,Bahn2001}.

Transmission electron microscopy~\cite{Rodrigues2000} shows that in the last stage of 
the stretching process, nanocontacts of Au are crystalline and atom rearrangements take
place in such a way that one of the crystal directions [100], [110] or [111] lies in 
stretching direction of the wire, independent of the starting crystal orientation.
For that reason our starting configuration is a perfect fcc-lattice of 112 atoms in [100] 
direction (z direction) of length 2.65 nm with a cross-section of eight atoms. This central
wire is attached at both ends to two slabs of 288 atoms each, which are kept fixed. We 
also performed simulations with starting configurations of perfect lattices in [110] 
(central wire: 108 atoms; each slab: 196 atoms) and in [111] direction (central wire: 
240 atoms; each slab: 720 atoms). The stretching process is simulated by separating both
slabs symmetrically at fixed distance in every time step. The resulting stretching velocity
of 2 m/s is much bigger than in the experiments, but it is small compared with the speed of
sound in the material (3000 m/s). Thus the wire can re-equilibrate between successive 
instabilities, although collective relaxation processes might be 
suppressed~\cite{Todorov1993,Todorov1996}. The Newtonian equations of motion of 
the wire atoms are integrated via the velocity Verlet algorithm~\cite{Frenkel1996} with a time
step of 1.4~fs. A Nos\'e-Hoover thermostat~\cite{Frenkel1996} maintains the average temperature
at 4.2 K. We used periodic boundary conditions for the slabs in z direction and minimum
image convention for the slabs perpendicular to the z direction~\cite{Allen1987}. Before 
the stretching process, every atom of the wire gets a randomly chosen velocity (that is 
why every nanocontact evolves a little bit different from stretching to stretching) and
the wire is equilibrated for about 0.7~ns with periodic boundary conditions perpendicular 
to the z direction.

In order to test whether the conductance changes are correlated with atom rearrangements
in the nanocontact, we calculate the radius of the minimum cross-section perpendicular
to the stretching direction as defined in Ref.~[\onlinecite{Bratkovsky1995}]. For this 
purpose, every atom is represented by a sphere with the volume of the 
elementary cell in the fcc-lattice. For a given configuration a slice with a width of 
about the interlayer distance~\cite{WidthSlice} is taken perpendicular to the stretching 
direction. Out of the volume of the atomic spheres overlapping with the slice, the radius 
of a cylinder which would fill the same volume in that slice is computed. The procedure is
repeated along the whole nanocontact and the smallest radius is taken as the radius 
of the minimum cross-section of a given configuration.

Moreover, during the stretching process, every 1.4~ps a configuration is recorded and 
the strain force of the nanocontact is computed following Finbow \emph{et 
al.}~\cite{Finbow1997}. Every 5.6~ps the corresponding conductance is calculated using
the method described in the next subsection.

\subsection{Conductance calculations}

The computation of conductance in atomic systems of the size that we consider here, is 
out of the scope of present ab-initio methods. For this reason, we chose to use a 
tight-binding model based on the parameterization introduced in Ref.~[\onlinecite{Mehl1996}].
As mentioned in the introduction, the tight-binding models have been very successful
in the description of the transport of metallic atomic contacts~\cite{Cuevas1998a,Cuevas1998b}. 
Our approach follows closely the one of Refs.~[\onlinecite{Cuevas1998a,Cuevas1998b}],
with the only difference that in this case we use a non-orthogonal basis, which 
introduces some minor changes as explained in Refs.~[\onlinecite{Emberly1998,Brandbyge1999}].

The first step in our approach is the description of the electronic structure of the
atomic contacts, which is done in terms of the following tight-binding Hamiltonian 
written in a non-orthogonal basis

\begin{equation}
\hat{H} =  \sum_{i \alpha, \sigma} \epsilon_{i \alpha}
c^{\dagger}_{i \alpha \sigma} c_{i \alpha \sigma}
+ \sum_{i \alpha \neq j \beta, \sigma} v_{i\alpha,j\beta}
c^{\dagger}_{i \alpha,\sigma}
c_{j \beta,\sigma} ,
\label{Hamiltonian}
\end{equation}

\noindent
where $i,j$ run over the atomic sites and $\alpha,\beta$ denote the different
atomic orbitals, $\epsilon_{i \alpha}$ are the on-site energies and
$v_{i\alpha, j\beta}$ are the hopping elements. Additionally, we need the overlaps
between the different orbitals, $S_{i\alpha,j\beta}$. We take all these parameters
from the bulk parameterization of Ref.~[\onlinecite{Mehl1996}], which is
known to accurately reproduce the band structure of bulk materials.
The atomic basis is formed by the 9 orbitals: $5d,6s,6p$, which give
rise to the main bands around the Au Fermi energy. It is important to emphasize
that in this parameterization both, the hopping elements and the overlaps, are
functions of the distance between the atoms, which allows us to apply it
in combination with the MD simulations. These functions have a cut-off radius
that encloses up to five nearest neighbors in a bulk geometry.

In an atomic contact the local environment in the neck region is very
different to that of the bulk material. In particular, this fact can lead to
large deviations from the approximate local charge neutrality that typical metallic
elements exhibit. We correct this effect imposing local charge neutrality in all
the atoms of the nanowire through a self-consistent variation of the diagonal parameters
$\epsilon_{i \alpha}$.

The quantity that we are interested in is the low-temperature linear conductance
which can be expressed in terms of the Landauer formula

\begin{equation}
G = G_0 \mbox{Tr} \{ \hat{t}(E_F) \hat{t}^{\dagger}(E_F) \} = G_0 \sum_n T_n ,
\end{equation}

\noindent
where $\hat t$ is the transmission matrix of the contact, and the $T_n$'s are the
transmission eigenvalues at the Fermi energy $E_F$. The transmission matrix
can be calculated in terms of the Green functions of the contact as follows

\begin{equation}
\hat{t}(E) = 2 \; \hat{\Gamma}^{1/2}_L(E) \hat{G}^r_{C}(E) 
\hat{\Gamma}^{1/2}_R(E) .
\end{equation}

\noindent
Here, $\hat{\Gamma}_{L,R}$ are the scattering rate matrices given by 
$\hat{\Gamma}_{L,R} = \mbox{Im} ( \hat{\Sigma}_{L,R} )$, where $\hat{\Sigma}_{L,R}$
are the self-energies which contain the information of the electronic structure of
the leads and their coupling to the central part of the contact. In our case the 
leads start in the slabs with 288 atoms and the central part, denoted by $C$, is 
the central wire.  The self-energies can be expressed as 

\begin{equation}
\hat{\Sigma}_{L,R} = \left( \hat{v}_{CL,R} - E \hat{S}_{CL,R} \right)
\hat g_{L,R} \left( \hat{v}_{L,RC} - E \hat{S}_{L,RC} \right) , 
\end{equation}

\noindent
where $\hat{v}$ is the hopping matrix which describes the connection between the
central cluster and the leads, and $\hat S$ is the corresponding overlap matrix. 
In this expression, $\hat g_{L,R}$ are the Green functions of the uncoupled leads, 
which we assume to be the bulk Green functions of Au. Finally, the central cluster
Green functions, $\hat G_C$, are given by

\begin{equation}
\hat{G}_C(E) = \left[ E \hat{S}_C - \hat{H}_C -
\hat{\Sigma}_L (E) - \hat{\Sigma}_R (E) \right]^{-1} ,
\end{equation}

\noindent
where $\hat{H}_C$ and $\hat{S}_C$ are the Hamiltonian and the overlap matrix
of the central part, respectively.

\section{Examples of the contact formation}

In this section we discuss in detail some of the typical features that we observe
in single realizations of the Au contacts. In particular, we discuss the formation
of both single-atom contacts and chains.

\subsection{Single-atom contacts}

\begin{figure}[t]
\begin{center}
\includegraphics[width=\columnwidth,clip]{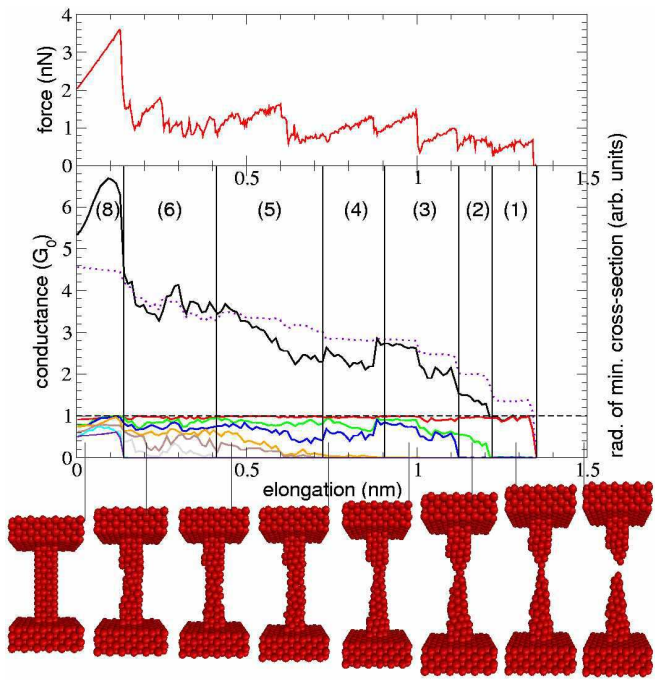}
\caption{\label{nochain1}(Color online). Formation of a dimer configuration of Au in the [100] 
direction at $T=4.2~K$. The upper panel show the strain forces as a function of the 
elongation of the contact. In the lower panel we show the conductance (solid line), 
the radius of the minimum cross-section (dashed line), and the channel transmissions
(lines in the bottom of the graph). The vertical lines define regions with different number of 
open channels ranging from 8 to 1. Below the graph we show snapshots of the breaking of
the contact.} 
\end{center}
\end{figure}

In Fig.~\ref{nochain1} we show the formation of what we could call a single-atom contact. 
To give a complete information of this process we have depicted the forces, the conductance,
the radius of the minimum cross-section, and the transmissions of the different conduction 
channels. In this example the conductance evolves during the elongation process from a value 
close to $5~G_0$ for the starting point to $1~G_0$ in the last stages before breaking~\cite{initial}.
The decrease of the conductance follows closely the evolution of the minimum cross-section. 
This correlation is particularly pronounced in the last stages when the contact has one or
two atoms in the narrowest part. However, it is worth stressing that the conductance does
not always follow the minimum cross-section, as can be seen in the elongation region between
0.75 and 1.0 nm. Sometimes there appear jumps in the conductance in regions where the minimum 
cross-section evolves smoothly. We interpret this fact as rearrangements of atoms away 
from the narrowest part of the wire. This interpretation is confirmed by the analysis of
the forces, as we explain in the next paragraph.

As shown in the upper panel of Fig.~\ref{nochain1}, one can see in the forces a 
series of regions where they increase linearly (elastic stages) and regions with
abrupt jumps (plastic stages). The first ones correspond to situations where the
structure remains fundamentally unchanged, and the latter correspond to the breaking
of bonds and the subsequent sudden atomic rearrangements. As commented in the 
introduction, this evolution of the forces has been measured for Au contacts at room 
temperature with the help of an atomic-force microscope~\cite{Rubio1996}.
The order of magnitude of the forces in Fig.~\ref{nochain1} is in good agreement
with this experiment.  Notice that also in the cases where the jumps in the conductance
are not correlated to abrupt changes in the minimum cross-section, one sees jumps in
the forces. This indicates that plastic deformations in regions away from the 
narrowest part of the wire have also an influence on the conductance. Thus,
we see that in the determination of the conductance, the minimum cross-section
is an important ingredient, but it is by no means the only one. This fact will
be crucial later on for the interpretation of the conductance histograms.

Another interesting feature in Fig.~\ref{nochain1} is the last conductance plateau.
It corresponds to a contact with the cross-section of one atom, but it is formed by 
a dimer of gold atoms. This dimer configuration is the most common geometry that
we find in the last stages of the contact breaking.
This type of configuration has been reported recently in MD simulations of
the breaking of Al wires~\cite{Jelinek2003}. Notice also that in the last plateau,
which is marked by a linear increase of the forces, the conductance is close to 
$1~G_0$ and it is clearly dominated by a single conduction channel. As explained 
in Refs.~[\onlinecite{Cuevas1998a,Cuevas1998b,Scheer1998}], this is due to the fact that the
number of channels is controlled by the number of valence orbitals in the
narrowest part of the contact. In the case of Au the main contribution to the 
density of states at the Fermi energy comes from the $6s$ band, i.e. there is
a single valence orbital. This implies that in the dimer configuration of
Fig.~\ref{nochain1}, there is only one possible path (conduction channel) which 
proceeds through the $s$ orbitals. It is also important to emphasize that during
the elongation of the contact we observe that the channels disappear one by one
in agreement with the experimental results reported in Ref.~[\onlinecite{Scheer1998}]. 
This will be discussed in more detail below. Let us point out that we consider a channel
being closed if its transmission is below $0.01~G_{0}$.  We attribute the successive 
closing of the channels to the fact that in the stretching process the Au atoms leave the
narrowest part one by one, and every Au atom contributes to the transport with one 
orbital~\cite{8channels}, which in turn can give rise to one conduction channel. 

\subsection{Atomic chains}

\begin{figure}[t]
\begin{center}
\includegraphics[width=\columnwidth,clip]{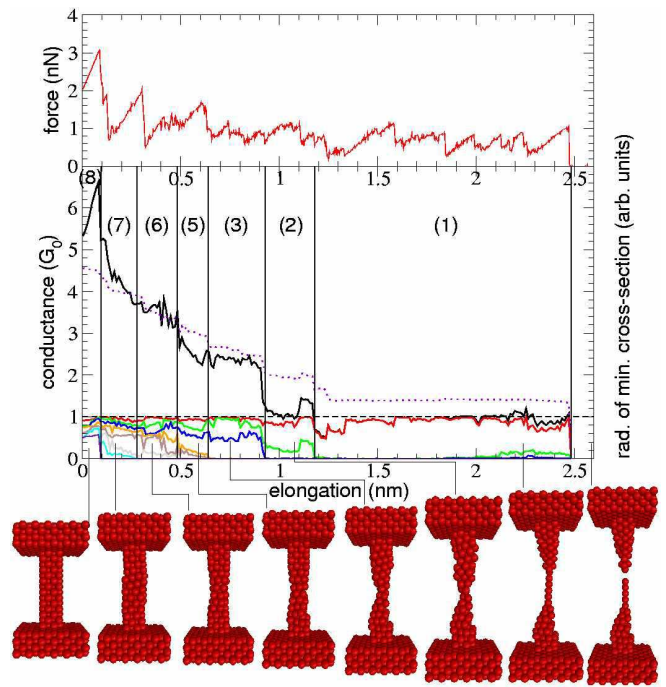}
\caption{\label{chain3}(Color online). Formation of Au chain in the [100] direction at $T=4.2~K$. 
The upper panel show the strain forces as a function of the elongation of the contact.
In the lower panel we show the conductance (solid line), the radius of the minimum 
cross-section (dashed line), and the channel transmissions (lines in the bottom of 
the graph). The vertical lines define regions with different number of open channels
ranging from 8 to 1. Below the graph we show snapshots of the breaking of the contact.} 
\end{center}
\end{figure}

In some occasions we observe that the contact does not break as in the configuration of
Fig.~\ref{nochain1}, but continues to form chains of several atoms. This is illustrated
in Fig.~\ref{chain3}. Experimental evidence of the formation of gold chains was
first reported independently by two groups~\cite{Ohnishi1998,Yanson1998}. The 
formation of chains had been already suggested by different 
simulations~\cite{Brandbyge1997,Sorensen1998,Finbow1997}. In the last years
there has been an intense activity in this topic. Thus for instance, there are now
several experiments confirming the existence of gold 
chains~\cite{Kizuka1998,Rodrigues2001,Takai2001}; chains have been also observed
in other materials~\cite{Smit2001}, and the forces during the formation of a Au chain
have been measured~\cite{Rubio2002}. From the theoretical side, many authors have analyzed
the formation, stability and conductance of gold chains~\cite{Ugarte2004,tbdftmd,Brandbyge1999,Torres1999,
Okamoto1999,Sanchez1999,Hakkinen1999,Emberly1999,Maria2000,Hakkinen2000,Bahn2001,Silva2001,
Palacios2002,Lee2004}. For a more complete list of references, see section
XI in Ref.~[\onlinecite{Agrait2003}]. We want to point out that so far there is
no theoretical analysis of the conductance during the chain formation available.

\begin{figure}[t]
\begin{center}
\includegraphics[width=\columnwidth,clip]{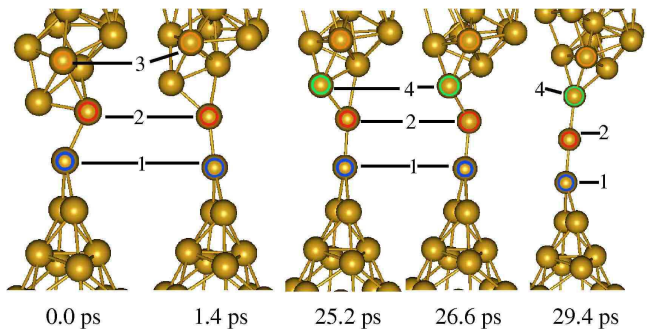}
\caption{\label{chainformation1}(Color online). Snapshots of the chain formation of 
Fig.~\ref{chain3}. At an initial time 0.0~ps the atom number 1 has three nearest neighbors
and the atom 2 has four neighbors. At 1.4~ps the bond between atom 2 and atom 3 breaks. At 
25.2~ps the atom 2 is only connected to two atoms of the electrode. At 26.6~ps 
another bond breaks and atom 4 is now in a similar environment as atom 2 at the 
beginning.}
\end{center}
\end{figure}

In our simulations for the [100] direction we have observed 11 chains out of 50 
stretching processes with different lengths: nine chains ranging from 3 atoms to 6 
atoms and three chains of 10, 12 and
14 atoms. Experimentally chains up to 8 Au atoms have been reported~\cite{Yanson1998}.
The mechanism of the chain formation has been explained by Bahn and Jacobsen\cite{Bahn2001}
in terms of many-body effects in metals. The main idea is that in certain metallic systems the 
binding energy per neighboring atom may increase as the number of neighbors decreases. Here, we 
want to illustrate how a chain is formed in real time. We show in Fig.~\ref{chainformation1}
five snapshots of the dynamics of the nanocontact of Fig.~\ref{chain3}.  At 0.0~ps an 
one-atom contact is shown with a typical dimer structure in the middle. The atom number
2 has four nearest neighbors and the atom 1 only three. Due to the many-body interactions
the bond between the atoms 2 and 3 breaks. After about 40~ps another bond breaks and
the atom 2 moves into the chain. The atom 4 is now in a similar situation as the atom 2 
was at the beginning, and the previous process can be repeated again.

Turning now to the transport properties of the chains, we see in Fig.~\ref{chain3} that
during the chain formation the conductance ranges from $0.6~G_0$ to $1.1~G_0$, exhibiting
a long flat plateau close to $1~G_0$. With respect to the
conductance value, this is in good agreement with
experiment~\cite{Smit2003}.
We observe that the conductance is mainly dominated by a single 
conduction channel. As in the case of a one-atom contact, the fact of having a single
dominant channel is a consequence of the fact that Au is a monovalent metal. At the end
of the stretching process, when the chain is made up of four or more atoms, we
typically observe the appearance of a second channel and sometimes even a third one
(see Fig.~\ref{chain3}). Our analysis shows that these additional channels are due to 
the contribution of the $d$ orbitals. With respect to the fluctuations of the conductance
during the formation of the chains, we want to stress that they are not an artifact of 
our conductance calculations, but they are clearly related to fluctuations in the
contact geometry. In Fig.~\ref{chain3} one can see that the structure in the 
conductance during the chain formation is accompanied by abrupt changes in the 
force values, i.e. by plastic deformations, which correspond to the incorporation
of another atom to the chain. We have calculated the conductance 
for all the configurations every 5.6~ps, including configurations that could
be unstable. In the experiment the stretching velocity is several orders of magnitude
smaller and such unstable configurations are averaged out. 

\section{Comparison with the experimental results}

\begin{figure}[t]
\begin{center}
\includegraphics[width=\columnwidth,clip]{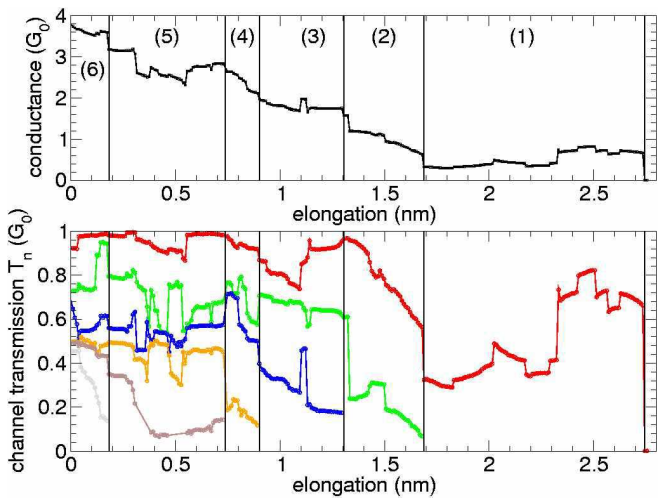}
\caption{\label{ScheerExp1}(Color online). Measured total conductance of a Au sample (top panel) and 
single-channel contributions (bottom panel) as a function of the electrode distance~\cite{ExpData}
(opening curve). The vertical lines define regions with different numbers of channels ranging 
from 6 to 1. The temperature was below $100$~mK. For experimental details see 
Refs.~[\onlinecite{Scheer2001},~\onlinecite{Scheer2002}].}
\end{center}
\end{figure}

In this section we show experimental results obtained following the technique of 
Refs.~[\onlinecite{Scheer1998},~\onlinecite{Scheer2001},~\onlinecite{Scheer2002}],
where part of the data has been already presented. In these experiments it is
possible to extract the full information of the individual transmission coefficients.
This is done inducing proximity superconductivity in a Au contact and analyzing the
super-conducting current-voltage characteristics~\cite{Averin1995,Cuevas1996}.
We refer the reader to Ref.~[\onlinecite{Scheer2002}] for further details.
Fig.~\ref{ScheerExp1} shows the conductance and the transmission coefficients of
a Au contact as a function of the elongation at 100~mK. We can see how the channels 
disappear one by one as the contact evolves. Notice that there is a very long plateau
of about 1~nm length, which corresponds to the formation of a chain. In this plateau 
the conductance changes between $0.3~G_0$ and $0.9~G_0$ and it is dominated by a single
channel. Notice also the abrupt changes of the conductance, which are most likely due
to incorporation of new atoms to the chain. 

\begin{figure}[t]
\begin{center}
\includegraphics[width=\columnwidth,clip]{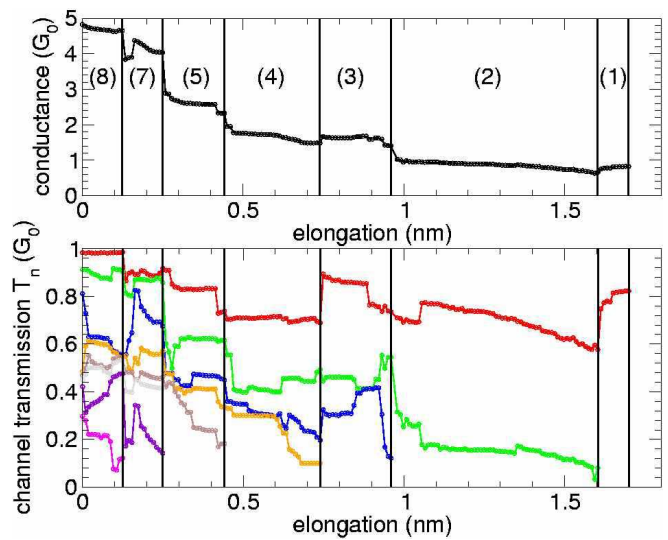}
\caption{\label{ScheerExp2}(Color online). Measured total conductance of a Au sample (top panel) and 
single-channel contributions (bottom panel) as a function of the electrode distance~\cite{ExpData}
(opening curve). The vertical lines define regions with different numbers of channels ranging 
from 8 to 1. The temperature was below $100$~mK. For experimental details see 
Refs.~[\onlinecite{Scheer2001},~\onlinecite{Scheer2002}].}
\end{center}
\end{figure}

\begin{figure}[t]
\begin{center}
\includegraphics[width=\columnwidth,clip]{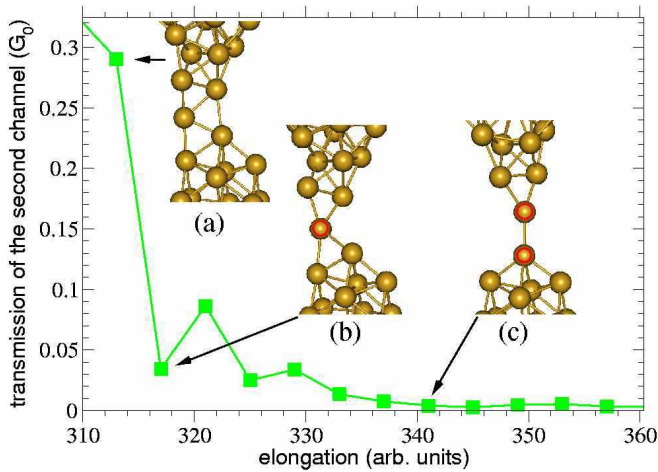}
\caption{\label{KonfChannel}(Color online). Transmission of the second channel during
the last stages of the breaking of a typical Au contact in the [100] direction (T=4.2K). 
The transmission of the first channel is about $0.7~G_{0}$.}
\end{center}
\end{figure}

In Fig.~\ref{ScheerExp2} we show another experimental conductance curve. As it can be
seen, there is a long plateau (about 0.7nm), where the conductance is close to $1~G_{0}$. 
Naively, one would expect the conductance to be dominated by a single conduction channel.
However, the analysis reveals the presence of two channels, and only in the very last
stages it reduces to one channel. In the two channel regime of the experiment in 
Fig.~\ref{ScheerExp2} the second channel has a transmission mainly below $0.2~G_{0}$ and
the total conductance is slightly below $1~G_{0}$. In our simulations we typically observe
two channels in the last stages for geometries in which there are two atoms in the
narrowest part displaced with respect to each other (see Fig.~\ref{KonfChannel}(a)). 
If there is only a single atom, the second channel gives only a small contribution to the
conductance (see Fig.~\ref{KonfChannel}(b)). 
As soon as we observe the dimer (see Fig.~\ref{KonfChannel}(c)), the conductance is 
largely dominated by a single channel. Our analysis of the character of the channels
suggests that for a single central atom like in Fig.~\ref{KonfChannel}(b) the second
channel is due to direct tunneling between the electrodes. 
In the dimer configuration the distance between the electrodes is considerably larger,
which leads to the disappearance of the second channel. The current then flows only through
the $s$ orbitals of the central atoms.

\begin{figure}[t]
\begin{center}
\includegraphics[width=\columnwidth,clip]{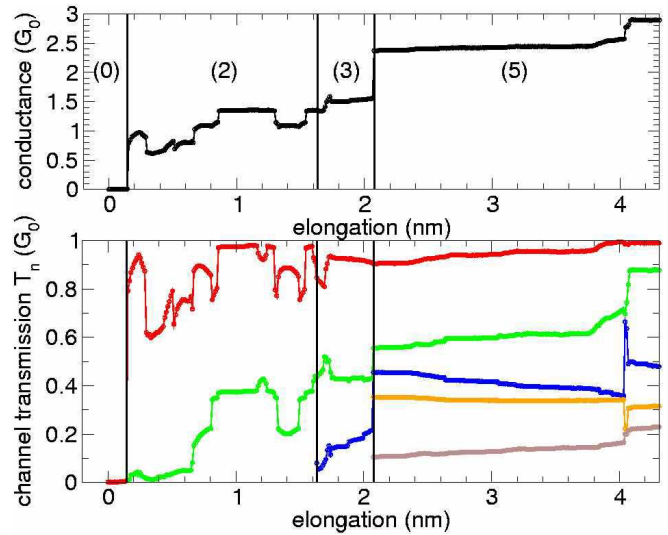}
\caption{\label{ScheerExp3}(Color online). Measured total conductance of a Au sample (top panel) and 
single-channel contributions (bottom panel) as a function of the electrode distance~\cite{ExpData}
(closing curve). The vertical lines define regions with different numbers of channels ranging 
from 2 to 5. The temperature was below $100$~mK. For experimental details see 
Refs.~[\onlinecite{Scheer2001},~\onlinecite{Scheer2002}].}
\end{center}
\end{figure}

The connection between the atomic structure and the number of channels described above
is supported by the experimental curve of Fig.~\ref{ScheerExp3} taken during the closing 
of a contact. In contrast to Fig.~\ref{ScheerExp1} and Fig.~\ref{ScheerExp2} and all 
the simulations, where the contacts were stretched (producing 
a so called opening curve), in Fig.~\ref{ScheerExp3} the two pieces of a nanocontact 
were brought into contact again after disruption (thereby producing a so called closing 
curve). It seems very unlikely to produce a dimer by reconnecting a teared nanocontact, 
since the atoms collapse back after rupture into the bulk. In fact, one expects a single 
atom coupled to the two pieces of the nanocontact as smallest contacts, i.e. one expects 
two channels at the beginning. Indeed, the typical closing curve (Fig.~\ref{ScheerExp3}) 
starts with two open channels.

\section{minimum cross-section histogram}

\begin{figure}[t]
\begin{center}
\includegraphics[width=0.8\columnwidth,clip]{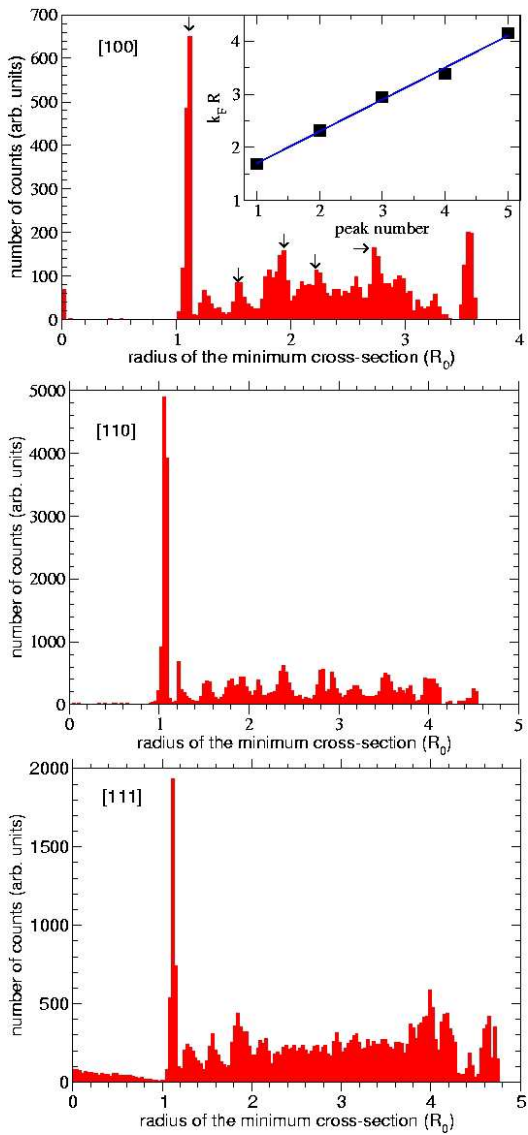}
\caption{\label{radiushistogram}(Color online). Histograms of the radius of the minimum 
cross-section $R$, which is normalized in units of the radius of an ideal
linear chain of gold atoms $R_0=1.268$ \AA. The temperature is 4.2K. The 
different panels correspond to different crystallographic orientations. 
The number of simulations used to construct the histograms is 50 for each
directions~\cite{HistoConvergence}. 
The inset in the upper panel shows the peak positions in the [100] histogram, 
converted to $k_F R$, as a function of the peak index. The arrows in the
upper panel indicate the peak positions used to construct this plot. The last
peak of the [100] histogram is artificially generated by the starting configurations.}
\end{center}
\end{figure}

In order to discuss in more detail the mechanical properties and stability of the 
contacts we have calculated histograms of the radius of the minimum cross-section for 
the three main crystallographic directions of the fcc-lattice.
The results are summarized in Fig.~\ref{radiushistogram}. In this figure
we have normalized the radius to the radius of an ideal linear chain $R_0 = 1.268$ \AA
~with nearest neighbor distance (${\mbox {lattice constant}}/\sqrt{2}$). The most 
prominent feature of the histograms is the presence of peaks, especially for the [100] 
direction. These peaks suggest the existence of exceptionally stable radii. This is
relatively surprising at low temperatures (4.2K), where the gold atoms do not have 
enough kinetic energy to explore many configurations in order to minimize the surface
energy. 

It is important to remark that these peaks generally do not appear at the values 
$\sqrt{n}\cdot R_{0}$ ($n=1,2,3,\ldots$), and therefore we cannot infer that
the cross-sections associated with these peaks correspond to multiple of the
cross-section of a single-atom contact. Furthermore in the three cases the peak
close to 1 is extremely pronounced. This is due to the fact that usually in the
last stages a dimer configuration or a chain of atoms is formed.

The next obvious question is whether this peak structure is an evidence of
shell effects as observed for alkali metals~\cite{Yanson1999} and suggested
more recently for gold contacts~\cite{Medina2003,Mares2004}. For shell
effects the peaks are expected to be equidistant when plotted as a function of 
the radius $R$ of the wire~\cite{Yanson1999}. To test this idea in the inset of
Fig.~\ref{radiushistogram}, we plot $k_FR$ as a function of the peak number for
the histogram of the [100] direction. Here, $k_F$ is the Fermi momentum, which
is assumed to take the free-electron value of $1.21 \times 10^8$ cm$^{-1}$. As 
one can see, the peak positions indeed follow a straight line with slope $0.60 
\pm 0.02$. This value of the slope lies in between the values observed in 
Ref.~[\onlinecite{Mares2004}] for electronic shells ($1.02$) and for atomic 
shells ($0.40$). From our analysis we cannot draw any conclusion on the appearance
of shell effects in our simulations for several reasons. First, the choice of the 
peak positions is not absolutely unambiguous, and second, a similar analysis for 
the histograms in the directions [110] and [111] does not show such a clear linear
relation between the radius and the peak position. It is worth stressing that even 
though the peak structure of these histograms cannot be easily interpreted in terms
of shell effects, it clearly shows that particularly stable configurations do exist,
which most likely correspond to meta-stable geometries which are formed at low temperatures.

\section{Conductance histogram}

\begin{figure}[t]
\begin{center}
\includegraphics[width=0.9\columnwidth,clip]{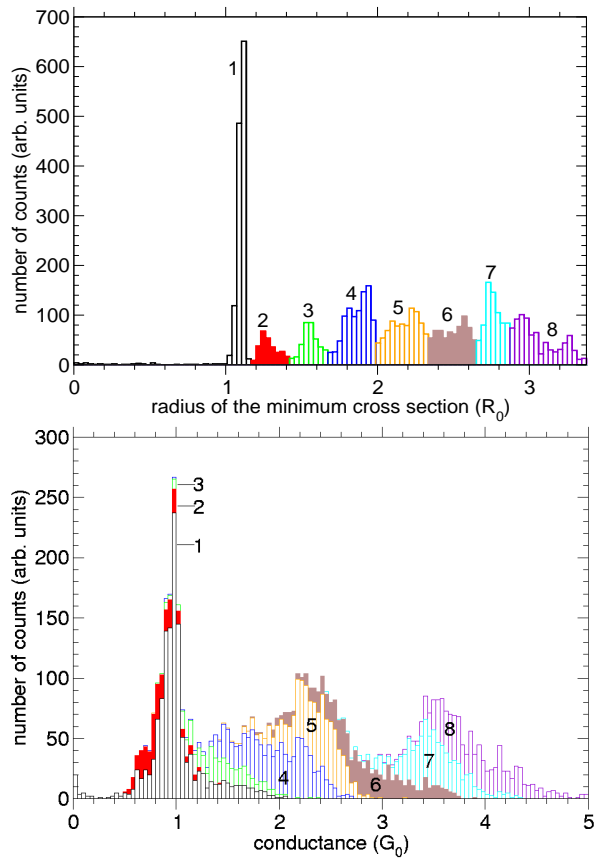}
\caption{\label{RhistxCondhistx1}(Color online). The lower panel shows the conductance histogram
for Au in [100] direction for $T=4.2~K$. We focus on the region $G < 5G_0$ after the initial
relaxation of the ideal contacts~\cite{initial}. The color/grey scales indicate how many counts
of a certain conductance value correspond to a certain region of the histogram of the radius of 
the minimum cross-section. This latter histogram is shown in the upper panel with the different
regions defined~\cite{RegionDefinition} with different color/grey scales. In order to make
it easier to distinguish the different colors in grey scaled print the colors are numbered.}
\end{center}
\end{figure}

In this section we want to address the main issue of this work, namely the origin
of the peaks in the conductance histograms. Are these peaks simply due to the peaks
in the minimum cross-section histograms as suggested in Ref.~[\onlinecite{Hasmy2001}]?
In order to answer this question, we have collected the conductance 
calculations~\cite{HistoConvergence} for all our MD simulations in the [100] direction 
at 4.2K. The results are displayed in the bottom panel of Fig.~\ref{RhistxCondhistx1}. 
As one can see, the conductance histogram has a pronounced peak close to $1~G_0$ and
two further maxima above $2~G_0$ and $3~G_0$. As we show below, the peak close to $1~G_0$ 
is highly correlated with the first peak in the corresponding minimum cross-section 
histogram of Fig.~\ref{RhistxCondhistx1} (upper panel), and it is therefore a consequence
of formation of dimer configurations or chains of atoms. 
However, the other peaks are not reflected in the conductance histogram in contrast to 
the postulated correlation between these two types of histograms for Al~\cite{Hasmy2001}. 
We explain this
fact as follows. As we pointed out in section III, the minimum cross-section is not the
only ingredient that controls the conductance, and the geometry of the narrowest part 
and the disorder in the contact play also an important role. This means in practice that
contacts with different radii can have similar values of the conductance. In particular,
this is true for small contacts like the one we have analyzed in this work. Of course,
the situation could be different for larger contacts, where semi-classical arguments 
are believed to provide a good description.

Let us now illustrate the correlation between the conductance histogram
and the minimum cross-section histogram. For this purpose, we show in 
Fig.~\ref{RhistxCondhistx1} both histograms in a color/grey scale representation. The
color/grey scale of the counts shows how many counts of a certain conductance value corresponds
to a certain region of the radius histogram. Notice that the peak in the conductance
histogram close to $1~G_0$ is clearly due to the contribution of contacts with the
cross-section of an atom, i.e. single-atom contacts or chains of atoms. The other
minimum cross-section regions overlap in the conductance histogram.

\begin{figure}[t]
\begin{center}
\includegraphics[width=\columnwidth,clip]{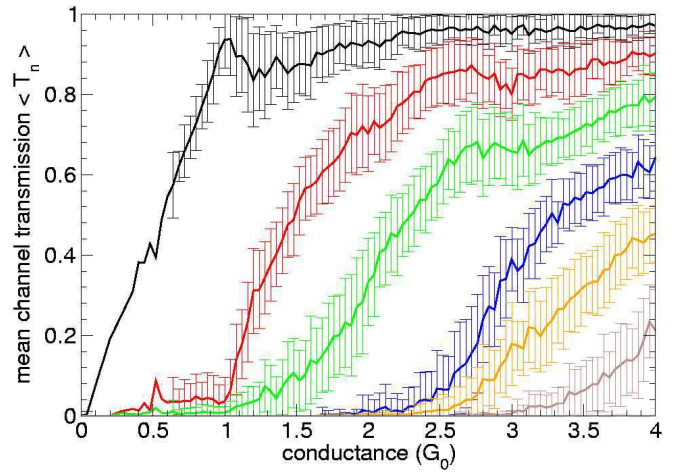}
\caption{\label{meanchannel1}(Color online). Mean value of the transmission coefficients
$\langle T_n \rangle$ ($n=1$ until $n=6$) as a function of the conductance in units of
$G_{0}$ for Au in [100] direction for $T=4.2~K$, i.e. corresponding to the histogram of 
Fig.~\ref{RhistxCondhistx1}. The error bars indicate the standard deviations
of the numerical results (if enough data points are available).}
\end{center}
\end{figure}

Another important piece of information can be obtained by analyzing the individual
transmission coefficients. Fig.~\ref{meanchannel1} shows the mean value \cite{MeanValue} of the
channel transmissions $<T_n>$ as a function of the total conductance for the conductance
histogram of Fig.~\ref{RhistxCondhistx1}. Notice that below $1~G_0$ the conductance is clearly
dominated by a single channel. In particular, for a total conductance of $1~G_{0}$
the second channel gives only a contribution about $0.036~G_{0}$. This is due to 
the fact that this region corresponds mainly to dimer configurations or chain of atoms,
where all the current proceeds through the $6s$ orbital of gold, as explained in 
section III. 

In the two channel regime of the experiment in Fig.~\ref{ScheerExp2} the second channel 
has a transmission mainly below $0.2~G_{0}$ and the total conductance is slightly below
$1~G_{0}$. Similar values are obtained in the simulations. For example, in 
Fig.~\ref{meanchannel1} we observe at a total mean conductance of about $1.1~G_{0}$, 
i.e. in the two channels regime (the third channel has only a mean transmission of under
$0.03~G_{0}$), a mean transmission of the second channel of about $0.18~G_{0}$. 
For higher conductance values, the channels are not completely open,
which explains the absence of conductance quantization in the histogram.

\section{Conclusions}

We have presented a comprehensive theoretical study of the mechanical and 
electrical properties of Au nano\-contacts at low temperatures and provide a 
comparison with experimental results on the transport channels. Our main goal
was to analyze how the interplay of these two type of properties is reflected
in the conductance histograms. We have shown that the histograms of the minimum
cross-section show a peak structure, which suggests that there are contact configurations
which are particularly stable. Semi-classical arguments, which are mainly based on 
the Sharvin formula or modifications of it, indicate that the conductance is
only controlled by the ratio between the radius of the contact and the Fermi 
wave length. With these arguments one would expect to see also a series of peaks
in the conductance histograms, which are correlated with the peaks of the histogram
of the minimum cross-section. However, in our calculations this is not the case, 
and only the peak close to $1~G_0$ is clearly correlated with the first peak of the 
histogram of the minimum cross-section. In other words, we find that (i) the minimum 
cross-section is not the only ingredient that determines the conductance and (ii)
the peak at $1~G_0$ is due to the formation of single-atom contacts and chains of atoms.
Our theoretical conductance histogram is in qualitative agreement
with the experimental findings at low temperatures. We note however, that we cannot
expect more than qualitative agreements due to the different stretching process
realization in the experiments and in the simulation. 

We have also found the following interesting results. In our simulations the 
most typical geometry in the last stages of the breaking of the contact is a
dimer configuration where two Au atoms are facing each other (see Fig.~\ref{nochain1}).
Our analysis of the conduction channels shows that the conductance is dominated
by a single channel in the last plateau, which corresponds to a single-atom contact or a chain of atoms. Moreover, the channels disappear one by one
as the section of the contact is reduced. Both observations are due to the fact 
that Au is a monovalent metal, where every atom contributes with a single valence
orbital to the transport. These finding are in good agreement with the experimental
observations (see Fig.~\ref{ScheerExp1}-\ref{ScheerExp2}). Furthermore the simulations
indicate a small contribution of the second channel to the total conductance if only 
one single atom is coupled to the electrodes. This finding can explain the number of
channels of the opening curve of the experiment (see Fig.~\ref{ScheerExp3}). 
In addition, as shown in Fig.~\ref{ScheerExp2} 
and in agreement with our simulations, the last conductance plateau in the experiment is 
not always carried by only one single channel, as is often assumed. 

Of course, it would be desirable to extend our analysis to higher temperatures
to investigate the appearance of shell effects~\cite{Medina2003,Mares2004}. On the 
other hand, it would be interesting to compare the results for Au with results
for other materials, in particular multivalent ones, to explain the differences
in the conductance histograms. Both type of studies are currently under progress.

\section{Acknowledgment}

It is a pleasure to thank M. H\"afner for useful discussions.
Support by the SFB~513 and granting of computer time from the NIC and the HLRS is
gratefully acknowledged. FP and JCC were financially supported by
the Landesstiftung Baden-W\"urttemberg within the ``Kompetenznetz
Funktionelle Nanostrukturen" and by the Helmholtz Gemeinschaft within the 
Nachwuchsgruppe program (contract VH-NG-029). The measurements have partially been
performed at CEA Saclay, France.
PN thanks the institutes for theoretical physics of the FU Berlin and the U Mainz for the hospitality during his sabbatical stay.



\begin{thebibliography}{}
\bibitem{Agrait2003}
N. Agra\"{\i}t, A. L. Yeyati and J.M. van Ruitenbeek, Phys. Rep. {\bf 377}, 81 (2003).

\bibitem{Muller1992} 
C.J. Muller, J.M. van Ruitenbeek, and L.J. de Jongh, Phys. Rev. Lett. {\bf 69}, 140 (1992).

\bibitem{Agrait1993} 
N. Agra\"{\i}t, J.G. Rodrigo, and S. Vieira, Phys. Rev. B {\bf 47}, 12345 (1993).

\bibitem{CQ} B.J. van Wees, H. van Houten, C.W.J. Beenakker, J.G.  Williamson, 
L.P. Kouwenhoven, D. van der Marel, and C.T. Foxon, Phys.  Rev. Lett. {\bf 60}, 848 (1988);
D.A. Wharam, T.J. Thornton, R. Newbury, M. Pepper, H. Ahmed, J.E.F. Frost, D.G. Hasko, 
D.C. Peacock, D.A. Ritchie, and G.A.C.  Jones, J. Phys. C {\bf 21}, L209 (1988).

\bibitem{Olesen1995}
L. Olesen, E. L{\ae}gsgaard, I. Stensgaard, F. Besenbacher, J. Schi{\o}tz, P. Stoltze, 
K.W. Jacobsen, and J.K. N{\o}rskov, Phys. Rev. Lett. {\bf 74}, 2147 (1995).

\bibitem{Krans1995} 
J.M. Krans, J.M. van Ruitenbeek, V.V. Fisun, I.K. Yanson and L.J. de Jongh, 
Nature {\bf 375}, 767 (1995).

\bibitem{Gai1996}
Z. Gai, Yi He, H. Yu, and W.S. Yang, Phys. Rev. B {\bf 53}, 1042 (1996).

\bibitem{Scheer1997} 
E. Scheer, P. Joyez, D. Esteve, C. Urbina and M.H. Devoret, Phys. Rev. Lett. {\bf 78}, 
3535 (1997).

\bibitem{Yanson1997} 
A.I. Yanson and J.M. van Ruitenbeek, Phys. Rev. Lett. {\bf 79}, 2157 (1997).

\bibitem{Cuevas1998a} 
J.C. Cuevas, A. L. Yeyati and A. Mart\'{\i}n-Rodero, Phys. Rev. Lett. {\bf 80}, 1066 (1998).

\bibitem{Cuevas1998b} 
J.C. Cuevas, A. L. Yeyati, A. Mart\'{\i}n-Rodero, G.  Rubio, 
C. Untiedt and N. Agra\"{\i}t, Phys. Rev. Lett. {\bf 81}, 2990 (1998).

\bibitem{Scheer1998} 
E. Scheer, N. Agra\"{\i}t, J.C. Cuevas, A. L. Yeyati, B. Ludoph, A. Mart\'{\i}n-Rodero,
G. Rubio, J.M. van Ruitenbeek and C. Urbina, Nature {\bf 394}, 154 (1998).

\bibitem{Landman1990}
U. Landman, W.D. Luedtke, N.A. Burnham, and R.J. Colton, Science {\bf 248}, 454 (1990).

\bibitem{Sutton1990} 
A.P. Sutton and J. B. Pethica, J. Phys.: Condens. Matter {\bf 2}, 5317 (1990).

\bibitem{Todorov1993} 
T.N. Todorov and A.P. Sutton, Phys. Rev. Lett. {\bf 70}, 2138 (1993).

\bibitem{Rubio1996} 
G. Rubio, N. Agra\"{\i}t and S. Vieira, Phys. Rev. Lett. {\bf 76}, 2302 (1996).

\bibitem{Yanson1999}
A.~I. Yanson, I.~K. Yanson, and J.M. van Ruitenbeek, Nature {\bf 400}, 144 (1999);
Phys. Rev. Lett. {\bf 84}, 5832 (2000).

\bibitem{Yanson2001}
A.~I. Yanson, I.~K. Yanson, and J.M. van Ruitenbeek, Phys. Rev. Lett. {\bf 87}, 216805 (2001).

\bibitem{Stafford1997}
C.A. Stafford, D. Baeriswyl and J. B\"urki, Phys. Rev. Lett. {\bf 79}, 2863 (1997).

\bibitem{Ruitenbeek1997} 
J.M. van Ruitenbeek, M.H. Devoret, D. Esteve, and C. Urbina, Phys. Rev. B {\bf 56}, 12566 (1997).

\bibitem{Yannouleas1997} 
C. Yannouleas and U. Landman, J. Phys. Chem. B {\bf 101}, 5780 (1997).

\bibitem{Hasmy2001} 
A. Hasmy, E. Medina, and P. A. Serena, Phys. Rev. Lett. {\bf 86}, 5574 (2001).  

\bibitem{Medina2003}
E. Medina, M. D\'{\i}az, N. Le\'on, C. Guerrero,  A. Hasmy, P. A. Serena,
and J. L. Costa-Kr\"amer, Phys. Rev. Lett. {\bf 91}, 026802 (2003).

\bibitem{Mares2004}
A.I. Mares, A.F. Otte, L.G. Soukiassian, R.H.M. Smit, and J.M. van Ruitenbeek, 
Phys. Rev. B {\bf 70}, 073401 (2004).

\bibitem{Ugarte2004}
P.Z. Coura, S.B. Legoas, A.S. Moreira, F. Sato, V. Rodrigues, S.O. Dantas, D. Ugarte, 
and D.S. Galv{\~a}o, Nanoletters {\bf 4}, 1187 (2004).

\bibitem{tbdftmd}
E.Z. da Silva, F.D. Novaes, A.J.R. da Silva, and A. Fazzio, Phys. Rev. B {\bf 69}, 115411 (2004).

\bibitem{Torres1996}
J.A. Torres and J.J. S\'aenz, Phys. Rev. Lett. {\bf 77}, 2245 (1996);
T. L\'opez-Ciudad, A. Garcia-Martin, A. J. Caamano, J. J.  Saenz, Surf. Sci. {\bf 440}, L887 (1999);
J. B\"urki, C. A. Stafford, X. Zotos, D. Baeriswyl, Phys. Rev. B {\bf 60}, 5000 (1999). 
A. Garcia-Martin, M. del Valle, J.J. S\'aenz, J.L. Costa-Kr\"amer, and P.A. Serena, 
Phys. Rev. B {\bf 62}, 11139 (2000).

\bibitem{Bratkovsky1995} 
A.M. Bratkovsky, A.P. Sutton and T.N. Todorov, Phys. Rev. B {\bf 52}, 5036 (1995).

\bibitem{Todorov1996} 
T.N. Todorov and A.P. Sutton, Phys. Rev. B {\bf 54}, 14234 (1996).

\bibitem{Mehrez1997}
H. Mehrez and S. Ciraci, Phys. Rev. B {\bf 56}, 12632 (1997).

\bibitem{Brandbyge1997}
M. Brandbyge, M.R. S{\o}rensen and K.W. Jacobsen, Phys. Rev. B {\bf 56}, 14956 (1997).

\bibitem{Sorensen1998}
M. R. S{\o}rensen, M. Brandbyge, and K. W. Jacobsen, Phys. Rev. B {\bf 57}, 3283 (1998).  
\bibitem{Buldum1998}
A. Buldum, S. Ciraci, I. P. Batra, Phys. Rev. B {\bf 57}, 2468 (1998).
\bibitem{Nakamura1999}
A. Nakamura, M. Brandbyge, L. B. Hansen, and K. W. Jacobsen, Phys. Rev. Lett. {\bf 82}, 1538 (1999). 

\bibitem{Ugarte2000}
V. Rodrigues, T. Fuhrer, and D. Ugarte, Phys. Rev. Lett. {\bf 85}, 4124 (2000).

\bibitem{Ugarte2003}
L.G.C. Rego, A.R. Rocha, V. Rodrigues, and D. Ugarte, Phys. Rev. B {\bf 67}, 045412-1 (2003). 

\bibitem{Muller1996}
C.J. Muller, J.M. Krans, T.N. Todorov and M.A. Reed, Phys. Rev. B {\bf 53}, 1022 (1996).

\bibitem{Costa-Kramer1997}
J.L. Costa-Kr\"amer, Phys. Rev. B {\bf 55}, R4875 (1997).

\bibitem{Hansen1997}
K. Hansen, E. L{\ae}gsgaard, I. Stensgaard, and F. Besenbacher, Phys. Rev. B {\bf 56}, 2208 (1997).

\bibitem{Li1998}
C.Z. Li and N.J. Tao, Appl. Phys. Lett. {\bf 72}, 894 (1998).

\bibitem{Ludoph2000}
B. Ludoph and J.M. van Ruitenbeek, Phys. Rev. B {\bf 61}, 2273 (2000).

\bibitem{Scheer2001}
E. Scheer, W. Belzig, Y. Naveh, M. H. Devoret, D. Esteve, and C. Urbina,
Phys. Rev. Lett. {\bf 86}, 284 (2001).

\bibitem{Scheer2002}
E. Scheer, W. Belzig, Y. Naveh, and C. Urbina, Adv. in Solid State Phys. {\bf 42}, 107 (2002).

\bibitem{Rubio2003}
G. Rubio-Bollinger, C. de las Heras, E. Bascones, N. Agra\"{\i}t, F. Guinea, 
and S. Vieira, Phys. Rev. B {\bf 67}, 121407 (2003).

\bibitem{Ohnishi1998}
H. Ohnishi, Y. Kondo and K. Takayanagi, Nature (London) {\bf 395}, 780 (1998).

\bibitem{Yanson1998}
A.I. Yanson, G. Rubio-Bolinger, H.E. van den Brom, N. Agra\"{\i}t, and
J.M. van Ruitenbeek, Nature (London) {\bf 395}, 783 (1998).

\bibitem{numEffort}
A typical computation of the structural evolution requires a computation
of $1.1\cdot 10^6$ MD steps, where after the equilibration of the system  
every $4\cdot 10^3$ MD steps the conductance is calculated. The resulting numerical
effort is about 400 CPU-hours (Intel Xeon CPU 2.80~GHz) for a single stretching process. 
The total CPU-time of our studies was about $20\cdot10^3$ CPU-hours.

\bibitem{Jacobsen1996} 
K.W. Jacobsen, P. Stoltze, J.K. N\o rskov, Surf. Sci. {\bf 366}, 394 (1996).

\bibitem{Stoltze1994} 
P. Stoltze, J. Phys.: Condens. Matter {\bf 6}, 9495 (1994).

\bibitem{Rodrigues2000} 
V. Rodrigues, T. Fuhrer, D. Ugarte, Phys. Rev. Lett. {\bf 85}, 4124 (2000).

\bibitem{Frenkel1996}
D. Frenkel and B. Smit, {\em Understanding Molecular Simulation}, 
Academic Press, 1996 and references therein.

\bibitem{Allen1987}
M.P. Allen and D.J. Tildesley, {\em Computer Simulation of Liquids},
Oxford Science Publications, 1987.

\bibitem{WidthSlice}
The width of the slice is $a/2$ for the [100] direction, where $a$ is the lattice constant,
$a/(2 \sqrt{2})$ for the [110] direction, and $a/\sqrt{3}$ for the [111] direction.

\bibitem{Finbow1997} 
G.M. Finbow, R.M. Lynden-Bell, I.R. McDonald, Molecular Physics {\bf 92}, 705 (1997).

\bibitem{Mehl1996}
M.J. Mehl and D.A. Papaconstantopoulos, Phys. Rev. B {\bf 54}, 4519 (1996).

\bibitem{Emberly1998}
E. Emberly and G. Kirczenow, Phys. Rev. Lett. {\bf 81}, 5205 (1998). 

\bibitem{Brandbyge1999}
M. Brandbyge, N. Kobayashi, and M. Tsukada, Phys. Rev. B {\bf 60}, 17064 (1999).

\bibitem{initial}
We observe in all the simulations an increase of the conductance during the first stages of
the stretching process, which is due to relaxation of the ideal fcc starting geometry. 
Only after this relaxation, an evolution characteristic of every single
contact begins.

\bibitem{Jelinek2003}
P. Jel\'{\i}nek, R. P\'erez, J. Ortega, and F. Flores, Phys. Rev. B {\bf 68}, 085403 (2003).

\bibitem{8channels}
For the starting cross-section with eight atoms one observes eight channels.

\bibitem{Kizuka1998}
T. Kizuka, Phys. Rev. Lett. {\bf 81}, 4448 (1998).

\bibitem{Rodrigues2001}
V. Rodrigues and D. Ugarte, Phys. Rev. B {\bf 63}, 073405 (2001).

\bibitem{Takai2001}
Y. Takai, T. Kawasaki, Y. Kimura, T. Ikuta, and R. Shimizu, Phys. Rev. Lett. 
{\bf 87}, 106105 (2001).

\bibitem{Smit2001}
R.H.M. Smit, C. Untiedt, A.I. Yanson, and J.M. van Ruitenbeek
Phys. Rev. Lett. {\bf 87}, 266102 (2001).

\bibitem{Rubio2002}
G. Rubio-Bollinger, S. R. Bahn, N. Agra\"{\i}t, K. W. Jacobsen, and S. Vieira
Phys. Rev. Lett. {\bf 87}, 026101 (2001).

\bibitem{Torres1999}
J.A. Torres, E. Tosatti, A. Dal Corso, F. Ercolessi, J.J. Kohanoff, F.D. Di Tolla,
and J.M. Soler, Surf. Sci. Lett. {\bf 426}, L441 (1999).

\bibitem{Sanchez1999}
D. S\'anchez-Portal, E. Artacho, J. Junquera, P. Ordej\'on, A. Garc\'{\i}a, and J. M. Soler
Phys. Rev. Lett. {\bf 83}, 3884 (1999).

\bibitem{Okamoto1999}
M. Okamoto and K. Takayanagi , Phys. Rev. B {\bf 60}, 7808 (1999).

\bibitem{Hakkinen1999}
H. H\"akkinen, R.N. Barnett, and U. Landman , J. Phys. Chem. B {\bf 103}, 8814 (1999).

\bibitem{Emberly1999}
E.G. Emberly and G. Kirczenow, Phys. Rev. B {\bf 60}, 6028 (1999).

\bibitem{Maria2000}
L. De Maria and M. Springborg, Chem. Phys. Lett. {\bf 323}, 293 (2000).

\bibitem{Hakkinen2000}
H. H\"akkinen, R.N. Barnett, A.G. Scherbakov, and U. Landman, J. Phys. Chem. B 
{\bf 104}, 9063 (2000).

\bibitem{Silva2001}
E.Z. da Silva, A.J.R. da Silva, and A. Fazzio, Phys. Rev. Lett. {\bf 87}, 256102 (2001).

\bibitem{Palacios2002}
J.J. Palacios, A.J. P\'erez-Jim\'enez, E. Louis, E. SanFabi\'an, and J. A. Verg\'es,
Phys. Rev. B 66, 035322 (2002).

\bibitem{Lee2004}
Y.J. Lee, M. Brandbyge, M.J. Puska, J. Taylor, K. Stokbro, and R.M. Nieminen,
Phys. Rev. B {\bf 69}, 125409 (2004).

\bibitem{Bahn2001}
S. R. Bahn and K.W. Jacobsen, Phys. Rev. Lett. {\bf 87}, 266101 (2001).

\bibitem{Smit2003}
R.H.M. Smit, C. Untiedt, G. Rubio-Bollinger, R. C. Segers, and J. M. van Ruitenbeek
Phys. Rev. Lett. {\bf 91}, 076805 (2003).

\bibitem{Averin1995}
D.~Averin and A.~Bardas, Phys. Rev. Lett. {\bf 75}, 1831 (1995).

\bibitem{Cuevas1996}
J.C.~Cuevas, A.~Mart\'{\i}n-Rodero and A. L. Yeyati, Phys. Rev. B {\bf 54}, 7366 (1996).

\bibitem{ExpData}
The absolute value of the electrode distance is arbitrary, only relative distances could 
be measured. In the bottom panel all the highest measured transmissions (circles) of 
each electrode distance are connected with straight lines, then the second highest 
transmissions of each electrode distance are connected and so on. This mapping of single
contributions of the transmission to certain channels cannot be measured experimentally.

\bibitem{HistoConvergence}
The convergence of the minimum cross-section histograms and the conductance histogram was
checked by comparing the normalized histograms constructed with 10, 20, 30, 40 and 50 
stretching processes. 

\bibitem{MeanValue}
For a certain value of the total conductance the channel contribution of the {\it n}th channel
(up to $n=8$ channels were observed in the simulations) of all 50 stretching processes are
added and then divided by 50. This gives the mean value  $<T_n>$ for the considered total
conductance value.
 

\bibitem{RegionDefinition}
The number of the regions of the minimum cross-section histogram is a little bit arbitrary.
However, if one defines instead of eight regions for example 21 (smaller) regions, the 
qualitative conclusions remain the same. 

\end{thebibliography}
\end{document}